\documentclass[10pt,conference]{IEEEtran}
\IEEEoverridecommandlockouts
\usepackage{bm} % for better boldface typography
\usepackage[noadjust]{cite} % Improves presentation of citations, like [3--7].
\usepackage{amsmath,amssymb,amsfonts}
\usepackage{algorithmic}
\usepackage{multirow}
\usepackage{graphicx}
\usepackage{textcomp}
\usepackage{xcolor}
\usepackage[table]{xcolor}
\usepackage{tikz}
\usetikzlibrary{shapes.geometric, arrows.meta}
\usepackage{adjustbox}
\usepackage{pgfplots}
\usepackage{tikz}
\usepackage{bm}
\pgfplotsset{compat=1.18}   % or compat=newest
\def\BibTeX{{\rm B\kern-.05em{\sc i\kern-.025em b}\kern-.08em
    T\kern-.1667em\lower.7ex\hbox{E}\kern-.125emX}}
\usepackage{makecell}
\usepackage{nicefrac}
\newcommand{\Reals}{\mathbb{R}}      % real numbers
\newcommand{\Naturals}{\mathbb{N}}   % natural numbers 1,2,3,...
\newcommand{\vect}[1]{\mathbf{#1}} % vector
\newcommand{\vectg}[1]{\bm{#1}} % italic vector (works also for greek)
\renewcommand{\vect}[1]{\vectg{#1}} % switch by default to second version!
\newcommand{\set}[1]{\mathcal{#1}}           % set
\newcommand{\cset}[1]{\mathcal{#1}^{\tn{c}}} % complement set
\newcommand{\ecard}[1]{|#1|}

\newcommand{\mat}[1]{\mathsf{#1}} % matrices (capital greek symbols work too)
 
\newcommand{\trace}[2][]{\operatorname{tr}_{#1}\left(#2\right)}
\newcommand{\etrace}[2][]{\operatorname{tr}_{#1}(#2)} 
\newcommand{\bigtrace}[2][]{\operatorname{tr}_{#1}\bigl(#2\bigr)}

\newcommand{\eqdef}{\triangleq} % definition

\newcommand{\argmin}{\operatorname*{argmin}}

\newcommand{\enorm}[2][]{\|#2\|_{#1}}

\newcommand{\eket}[1]{\vert{#1}\rangle}

\newcommand{\tn}[1]{\textnormal{#1}}

\usepackage{soul} % hyphenation for letterspacing, underlining, and more
\usepackage[colorinlistoftodos, textsize=footnotesize]{todonotes} % add inline or margin comments

\newcommand{\blue}{\color{blue}}

\usepackage{balance}
\usepackage{hyperref}

\allowdisplaybreaks

\begin{document}
\bstctlcite{IEEEtran:default}

\title{Can Entanglement-Induced Feature-Map Designs Improve Projected Quantum Kernels?}

\author{\IEEEauthorblockN{Ravi Kumar Jha, Hsuan-Yin Lin, and Eirik Rosnes}
\IEEEauthorblockA{Simula UiB, N--5006 Bergen, Norway\\
Emails: \{ravi, lin, eirikrosnes\}@simula.no}}

\title{Can Entanglement-Induced Feature-Map Designs Improve Projected Quantum Kernels?\thanks{The authors acknowledge the financial support provided in part by the SURE-AI Centre Grant 357482, Research Council of Norway.}}

\maketitle
\begin{abstract}% \label{abstract}
Projected quantum kernels (PQKs) have emerged as a notable approach to circumvent the hardness of machine learning tasks arising from the exponential dimensionality of the quantum Hilbert space. However, recent PQK studies often rely heavily on extensive classical hyperparameter tuning of the radial basis function (RBF) kernel parameter and the support vector machine regularization parameter. The extensive tuning can mask the true representational power of the underlying PQKs induced by quantum feature maps, which has not been explored in detail. Thus, this work aims to expose the structural capabilities of PQKs by isolating and evaluating their native inductive bias through comparisons between untuned and tuned PQKs against the RBF baseline. Furthermore, we propose novel quantum feature maps that leverage two-qubit entanglement to capture multivariate correlations in IoT data. The results show that extensive tuning can lead to competitive performance architectures, while our proposed entanglement topologies exhibit a significantly stronger native inductive bias. This reveals that superior quantum architectural design can yield structurally better PQKs with improvements arising from quantum-mechanical properties rather than from classical hyperparameter tuning.
\end{abstract}

%\begin{IEEEkeywords}
%Projected Quantum Kernel, Quantum Feature Map, Hyperparameter Tuning, IoT data, Classification 
%\end{IEEEkeywords}

\section{Introduction}

Quantum computing (QC) for machine learning (ML), also known as quantum machine learning (QML), has emerged as a promising research direction for ML tasks~\cite{RebentrostMohseniLloyd14_1, Biamonte-etal17_1}. This development is primarily driven by the need for computationally efficient paradigms to address complex systems, data modeling and simulation, data structures and design patterns, and energy estimation. QML leverages intriguing quantum-mechanical phenomena, such as information processing using quantum bits (qubits), superposition, and entanglement~\cite{NielsenChung10_1}. Notable algorithms have been developed over the last decade to enhance existing ML capabilities by leveraging these quantum properties. In the vicinity of  ML, algorithms such as quantum neural networks, quantum principal component analysis (PCA), and quantum kernels have been developed to explore various advancements~\cite{RodriguezDiazGutierrez-AvilesTroncosoMartínez-Alvarez25_1}. This work focuses on quantum kernels. % NielsenChung10_1

The mathematical framework of quantum kernels is straightforward and aligns with kernel theory, in which information is processed by mapping it to state vectors living in a large quantum feature space~\cite{SchuldKilloran19_1, Schuld21_1sub}. The high-dimensional feature vectors need not be computed explicitly, and instead, a quantum device directly evaluates the inner products of quantum states. This overlap forms a quantum kernel, providing the similarity measure that drives kernel-based methods~\cite{SchuldKilloran19_1, Schuld21_1sub, RodriguezDiazGutierrez-AvilesTroncosoMartínez-Alvarez25_1}.

One ongoing challenge in QML is how to effectively encode classical inputs into quantum states through a \emph{quantum feature map}, or simply a \emph{feature map}~\cite{SchuldKilloran19_1, Schuld21_1sub, KhanAmanSikdar24_1}. In kernel methods, feature maps are crucial because kernel performance depends on how effectively they capture task-relevant structure through pairwise inner products of mapped inputs. A feature map typically consists of single- and two-qubit gates together with entangling  operations~\cite{SchuldKilloran19_1, Schuld21_1sub, NguyenChen22_1, KhanAmanSikdar24_1}. An early demonstration of quantum advantage was achieved using a quantum fidelity kernel on synthetic data, executed on an IBM quantum device. This work remains influential in introducing the \emph{$\mat{ZZ}$ feature map}~\cite{Havlicek-etal19_1}. This has inspired many useful formulations of quantum kernels in various applications, such as solving differential equations, hybrid neuro-inspired modeling, and data analysis~\cite{RebentrostMohseniLloyd14_1, LiuArunachalamTemme21_1, PaineElfvingKyriienko23_1, JhaKasabovBhattacharyyaCoylePrasad25_2}.

Standard quantum kernels assess data similarity by computing the global inner product between quantum states in a high-dimensional Hilbert space. However, as the number of qubits increases, these global states become increasingly orthogonal, causing the kernel values of data points to decay exponentially toward zero, called \emph{exponential concentration}~\cite{ThanasilpWangCerezoHolmes24_1}. Thus, exponential concentration severely hinders learning by diminishing the variance of the kernel matrix, making data points nearly indistinguishable and thereby limiting the model's predictive capability. To mitigate the exponential scaling of the Hilbert space, the \emph{projected quantum kernel} (PQK) was introduced~\cite{Huang-etal21_1}. Instead of relying on global state overlap, a PQK projects the quantum state back into a lower-dimensional classical feature space via local observable measurements, preserving relevant nonlinear quantum correlations while maintaining the discriminative power required for effective kernel estimation~\cite{Huang-etal21_1, SchnabelRoth25_1, D’Amore-etal26_1}. However, the PQK formulation appears to rely heavily on extensive classical hyperparameter tuning of the \emph{radial basis function (RBF)} kernel parameter $\gamma$ and the support vector machine (SVM) regularization parameter $\xi$. Therefore, this work critically examines current literature practices by addressing the following key objectives:
\begin{itemize}
\item To assess the extensive classical tuning influences on both PQKs and the RBF, we isolate their native inductive biases by comparing their tuned and untuned performances.
\item To introduce two novel entanglement-induced feature maps that achieve empirical gains while demonstrating reduced hyperparameter sensitivity over untuned PQKs.
\end{itemize}
%In standard PQK frameworks, the projected vectors are fed into a classical radial basis function (RBF) kernel to compute the final kernel matrix entries.
%\item The extent to which extensive tuning is actually necessary, and we demonstrate this by systematically isolating and evaluating the native inductive bias of PQKs.
%\item We propose two new variations of quantum feature maps that provided empirical gains with reduce sensitivity with the existing PQKs.

To this end, we evaluate the models under two distinct settings: untuned baselines and robustly tuned models. This framework ensures reliable assessment, aiming to provide a missing understanding of how structurally aligned quantum topologies can natively separate complex data and improve understanding of PQK models. %The content of this paper is as follows: Section II provides the PQK methodology and data details, Section III presents the results and discussion, and Section IV provides the conclusion and future work.%

\section{Preliminaries}
\label{sec:preliminaries}

\subsection{Notation}
\label{sec:notation}

$\Naturals$ and $\Reals$ denote the sets of positive integers and reals, respectively. Moreover, $[a]\eqdef\{1,2,\ldots,a\}$ for $a\in\Naturals$. Vectors are boldfaced, e.g., $\vect{x}$. Matrices and sets are represented by capital sans serif letters and calligraphic uppercase letters, respectively, e.g., $\mat{X}$ and $\set{X}$. The single-qubit rotation operators about the $x$-, $y$-, and $z$-axes are denoted by $\mat{R}_{\tn{X}}$, $\mat{R}_{\tn{Y}}$, and $\mat{R}_{\tn{Z}}$, respectively, and $\mat{H}$ is the Hadamard gate. We use $\rho$ to denote a density operator, and $\etrace{\mat{X}}$ and $\etrace[\set{A}]{\mat{X}}$ for the trace of $\mat{X}$ and the partial trace of $\mat{X}$ over a subsystem $\set{A}$, respectively. Tensor products of Pauli operators are written without the tensor-product symbol, e.g., $\mat{X}\mat{X}$ denotes $\mat{X}\otimes\mat{X}$. % $\rho_{\set{M}}(\vect{x})$ implies the reduced density matrix, $\set{S}$ denotes subsystem, and $\set{P}$ is nontrivial Pauli observables.

\subsection{Internet of Things Data}
\label{sec:datainfo}

\input{figs_QAI/pca.tex} % PCA plot

Internet of Things (IoT) platforms capture real-time environmental changes across a network of connected devices and sensors. The idea of using IoT data in this work is to extend the real-world applications of current quantum models beyond synthetic datasets, demonstrating how a quantum method handles complex, nonlinear correlations in sensor networks. Moreover, the inclusion of this data facilitates a fair comparison with ML models, as it does not require any feature reduction or selection method, given the moderate sample size and dimensionality, and it has previously been used in QML work~\cite{Khan-etal24_1, D’Amore-etal26_1}. The data is open access on the IoT platform COGITO~\cite{D’Amore-etal26_1}. The dataset contains $N=2865$ samples and the six features  \emph{illuminance}, \emph{blinds}, \emph{lamps}, \emph{relative humidity (rh)}, \emph{carbon dioxide (CO2)}, and \emph{temperature (temp)} for classifying room occupancy. Standard data pre-processing was used to remove irrelevant variables, such as levels and timestamps, and redundant records. The binary classification task of \emph{occupied vs nonoccupied} was normalized before being used in distinct PQKs and classical models. Fig.~\ref{fig:pca} visualizes the IoT data using the first two principal components obtained by PCA. The plot shows that the two classes are partially separable but remain noticeably overlapping, illustrating the nontrivial structure of the classification task.

\subsection{Data Encoding} 
\label{sec:encode}

Encoding classical inputs into quantum states, commonly referred to as \emph{data encoding}, is a critical step in QML. A data-encoding scheme that maps classical feature inputs into high-dimensional quantum states is known as a quantum feature map. In a basic formulation, a $d$-dimensional feature input $\vect{x}\in\Reals^{d}$ is encoded into an $n$-qubit quantum state $\eket{\phi({\vect{x}})}$ through a data-dependent unitary $\mat{U}({\vect{x}})$, namely, $\eket{\phi(\vect{x})}\eqdef\mat{U}(\vect{x})|0\rangle^{\otimes n}$. The corresponding density matrix is denoted by $\rho(\vect{x})$. This encoding via feature maps transforms the input data into a high-dimensional Hilbert space. Finding a suitable feature map is therefore a central component of quantum kernel models, where predictions can be made through kernel evaluations~\cite{SchuldKilloran19_1, Havlicek-etal19_1, JhaKasabovBhattacharyyaCoylePrasad25_2, ThanasilpWangCerezoHolmes24_1}. In this work, following the IoT setting in~\cite{D’Amore-etal26_1}, the number of qubits is chosen to be the same as the number of IoT input features, namely $n=6$. In general, however, the input dimension, the number of qubits, and the projected feature dimension need not coincide.
% We next introduce the PQK formulation.
% In this work, the number of qubits used is the same as the IoT feature dimensions, both six. In general, the input dimension and the number of qubits need not be the same. We discuss the PQK formulation next.

\subsection{Projected Quantum Kernel} 
\label{sec:pqkintro}

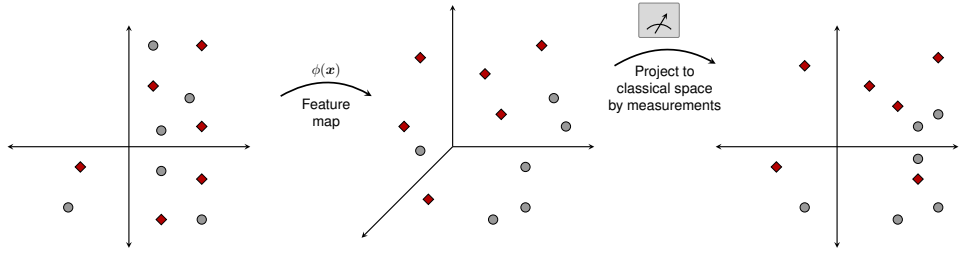
\begin{figure*}[!t]
\centering
\begin{adjustbox}{width=\textwidth,max totalheight=3.25cm,keepaspectratio}
\begin{tikzpicture}[
    data circle/.style={circle, draw=black, fill=gray!80, inner sep=2pt, minimum size=6.5pt},
    data diamond/.style={diamond, draw=black, fill=red!70!black, inner sep=1.5pt, minimum size=7.5pt},
    axis line/.style={thick, <->, >=stealth},
    axis line pos/.style={thick, ->, >=stealth},
    text label/.style={align=center, font=\sffamily}
]

% --- Left Plot (2D Input Space) ---
\begin{scope}[shift={(0,0)}]
    \draw[axis line] (-3, 0) -- (3, 0);
    \draw[axis line] (0, -2.5) -- (0, 3);
    % Data points
    \node[data circle] at (0.6, 2.5) {};
    \node[data diamond] at (1.8, 2.5) {};
    \node[data diamond] at (0.6, 1.5) {};
    \node[data circle] at (1.5, 1.2) {};
    \node[data circle] at (0.8, 0.4) {};
    \node[data diamond] at (1.8, 0.5) {};
    \node[data diamond] at (-1.2, -0.5) {};
    \node[data circle] at (0.8, -0.6) {};
    \node[data diamond] at (1.8, -0.8) {};
    \node[data circle] at (-1.5, -1.5) {};
    \node[data diamond] at (0.8, -1.8) {};
    \node[data circle] at (1.8, -1.8) {};
\end{scope}
% --- First Arrow (Feature Map) ---
\begin{scope}[shift={(3.5, 0)}]
    \draw[very thick, ->, >=stealth] (0.25, 1.25) to[bend left=30] 
        node[midway, above] {$\phi(\vect{x})$} 
        node[midway, below=7pt, text label] {Feature\\ map} (2.5, 1.25);
\end{scope}
% --- Middle Plot (3D Quantum Space) ---
\begin{scope}[shift={(8, 0)}]
    \draw[axis line pos] (0, 0) -- (0, 3.5);     % Y-axis (up)
    \draw[axis line pos] (0, 0) -- (3.5, 0);     % X-axis (right)
    \draw[axis line pos] (0, 0) -- (-2.25, -2.25); % Z-axis (coming out)
    % Elevated / separated points
    \node[data diamond] at (-0.8, 2.2) {};
    \node[data diamond] at (0.8, 1.8) {};
    \node[data diamond] at (2.2, 2.5) {};
    \node[data diamond] at (-1.2, 0.5) {};
    \node[data diamond] at (1.2, 0.8) {};
    \node[data diamond] at (-0.6, -1.3) {};
    \node[data circle] at (2.5, 1.2) {};
    \node[data circle] at (1.8, -0.5) {};
    \node[data circle] at (2.8, 0.5) {};
    \node[data circle] at (-0.8, -0.1) {};
    \node[data circle] at (1, -1.8) {};
    \node[data circle] at (1.8, -1.5) {};
\end{scope}
% --- Second Arrow (Projection) ---
\begin{scope}[shift={(12, 1.5)}]
    % Measurement Icon
    \begin{scope}[shift={(0.6, 1.2)}]
        \draw[fill=gray!30, draw=gray, rounded corners=1pt] (0,0) rectangle (1, 0.85);
        \draw[thick] (0.2, 0.2) to[bend left=45] (0.8, 0.2);
        \draw[thick, ->, >=stealth] (0.4, 0.1) -- (0.7, 0.6);
    \end{scope}
    \draw[very thick, ->, >=stealth] (0.0, 0.5) to[bend left=30] 
        node[midway, below=7pt, text label] {Project to\\ classical space\\ {by measurements}} (2.5, 0.55);
\end{scope}
% --- Right Plot (2D Projected Space) ---
\begin{scope}[shift={(17.5, 0)}]
    \draw[axis line] (-3, 0) -- (3, 0);
    \draw[axis line] (0, -2.5) -- (0, 3);
    % Better Separated points
    \node[data diamond] at (-0.8, 2) {};
    \node[data diamond] at (0.8, 1.5) {};
    \node[data diamond] at (2.5, 2.2) {};
    \node[data diamond] at (-1.5, -0.5) {};
    \node[data diamond] at (1.5, 1) {};
    \node[data diamond] at (2, -0.8) {};
    \node[data circle] at (2.5, 0.8) {};
    \node[data circle] at (2, 0.5) {};
    \node[data circle] at (2, -0.3) {};
    \node[data circle] at (-0.8, -1.5) {};
    \node[data circle] at (1.5, -1.8) {};
    \node[data circle] at (2.5, -1.5) {};
\end{scope}
\end{tikzpicture}
\end{adjustbox}
\caption{A schematic design of the PQK architecture, where the raw inputs are first mapped into quantum feature space using a quantum feature map $\phi$, and finally projected back into a classical space through measurement.}
\label{fig:pqk1}
\end{figure*}

PQKs are a class of quantum models that project quantum states into an approximate classical space~\cite{Huang-etal21_1}. In the PQK framework, rather than computing the fidelity between quantum states, the output quantum states are measured to extract the expectation values of Pauli observables, further discussed in Sec.~\ref{sec:methodology}. Fig.~\ref{fig:pqk1} shows a schematic design of the PQK formulation. The original idea of this formulation was to address the challenges posed by the exponential dimensionality of the quantum Hilbert space endured by the standard quantum kernel approach~\cite{Huang-etal21_1, ThanasilpWangCerezoHolmes24_1, SchnabelRoth25_1, D’Amore-etal26_1}. The PQK formulation demonstrated merit in computational efficiency, with linear scaling of the required QC resources. However, PQKs still utilize the quantum Hilbert space and therefore can still pose a hardness barrier to classical simulation. We use the PQK formulation from~\cite[eq.~(14)]{D’Amore-etal26_1} (see also~\cite[Supplementary information]{Huang-etal21_1}):
\begin{IEEEeqnarray*}{rCl}
    \IEEEeqnarraymulticol{3}{l}{%
      \kappa_{\tn{PQK}}(\vect{x},\vect{x}')
    }\nonumber\\*%
    & \eqdef &
    \exp\Biggl(-\gamma\sum_{\set{M}\in\set{S}}\sum_{\mat{O}\in\set{P}}
    \Bigl[\bigtrace{\mat{O}\,\rho_{\set{M}}(\vect{x})}-\bigtrace{\mat{O}\,\rho_{\set{M}}(\vect{x}')}\Bigr]^2
    \Biggr),\IEEEeqnarraynumspace\label{eq:def_pqk}
\end{IEEEeqnarray*}
where $\gamma>0$ is a kernel hyperparameter, $\set{S}$ denotes a collection of subsystems of the $n$-qubit system, and $\set{P}$ denotes a set of nontrivial Pauli observables whose locality is chosen consistently with the subsystems in $\set{S}$. For a subsystem $\set{M}$, $\rho_{\set{M}}(\vect{x})\eqdef\trace[\cset{\set{M}}]{\rho(\vect{x})}$ is the reduced density operator obtained by tracing out the complement subsystem $\cset{\set{M}}$.

% PQK illustration plot 

The PQK $\kappa_{\tn{PQK}}(\vect{x},\vect{x}')$ evaluates similarity % through the Euclidean distance between vectors of local
using the expectation values of selected observables computed from the reduced density matrices $\rho_{\set{M}}(\vect{x})$. It can therefore be interpreted as a classical RBF kernel applied to quantum-projected feature vectors. Different projection strategies can be used to extract these features from the quantum states.

In this work, we follow the same choices of subsystems and observable sets as in~\cite{D’Amore-etal26_1} to compute $\kappa_{\tn{PQK}}(\vect{x},\vect{x}')$ according to the locality level $m$. The measurement settings are as follows. (a) Single-qubit local measurements $(m=1)$: We set $\set{S}=\set{S}_1\eqdef\{\{1\},\{2\},\ldots,\{6\}\}$ and $\set{P}=\set{P}_1\eqdef\{\mat{X}, \mat{Y}, \mat{Z} \}$, yielding $18$ observables. (b) Two-qubit measurements on adjacent qubit pairs ($m=2$): We set $\set{S}=\set{S}_2\eqdef\{\{1,2\},\{2,3\},\{3,4\},\{4,5\},\{5,6\},\{6,1\}\}$ and $\set{P}=\set{P}_2\eqdef\{\mat{X}\mat{X}, \mat{Y}\mat{Y}, \mat{Z}\mat{Z}\}$, again yielding $18$ observables. (c) Combined single- and two-qubit measurements ($m \in \{1, 2\}$): We concatenate the projected features obtained using $\set{S}=\set{S}_1$ and $\set{P}=\set{P}_1$ for $m=1$, with those obtained using $\set{S}=\set{S}_2$ and $\set{P}=\set{P}_2$ for $m=2$, yielding $36$ observables~\cite{D’Amore-etal26_1}.

As a simple example for computing $\kappa_{\tn{PQK}}$, consider two input samples $\vect{x}_1$ and $\vect{x}_2$ encoded into density operators $\rho(\vect{x}_1)$ and $\rho(\vect{x}_2)$.
% (note that the description of the previous paragraph uses $n=6$).
We take $\set{S}=\set{S}'=\{\{1\},\{2\}\}$, and use the simple observable set $\set{P}=\set{P}'=\{\mat{Z}\}$ for illustration. % Assume two local observables $\mat{Z_1}$ and $\mat{Z_2}$ are measured, and their assumed expectation values are
Suppose that the corresponding expectation values are
\begin{IEEEeqnarray*}{rClCrCl}
    \etrace{\mat{Z}\rho_{\{1\}}(\vect{x}_{1})}& = & 0.8,
    &&\quad
    \etrace{\mat{Z}\rho_{\{2\}}(\vect{x}_{1})}& = & -0.2,
    \\
    \etrace{\mat{Z}\rho_{\{1\}}(\vect{x}_{2})}& = & 0.5,
    &&\quad
    \etrace{\mat{Z}\rho_{\{2\}}(\vect{x}_{2})}& = & 0.4.
\end{IEEEeqnarray*}
% \[
% \begin{aligned}
% \langle \mat{Z_1} \rangle_{\vect{x}_1} &= 0.8, &
% \langle \mat{Z_2} \rangle_{\vect{x}_1} &= -0.2,\\
% \langle \mat{Z_1} \rangle_{\vect{x}_2} &= 0.5, &
% \langle \mat{Z_2} \rangle_{\vect{x}_2} &= 0.4.
% \end{aligned}
% \]
% \textcolor{blue}{The measurements form the projected feature vectors $\vect{c}(\vect{x})$ as:
% \[
% \vect{c}(\vect{x})=
% \begin{bmatrix}
% \bigtrace{\mat{Z}_1\rho(\vect{x})}\\
% \bigtrace{\mat{Z}_2\rho(\vect{x})}
% \end{bmatrix}
% \!\implies\!
% \vect{c}(\vect{x}_1) \!=\!
% \begin{bmatrix}
% 0.8\\
% -0.2
% \end{bmatrix}\!, \,
% \vect{c}(\vect{x}_2) \!=\!
% \begin{bmatrix}
% 0.5\\
% 0.4
% \end{bmatrix}\!.
% \]}
% \textcolor{blue}{The squared Euclidean distance between these vectors is $\|vect{c}(\vect{x}_1)-\vect{c}(\vect{x}_2)\|^2 =0.45$.} 
Applying the PQK definition with $\gamma=1$, we obtain $\kappa_{\tn{PQK}}(\vect{x}_1,\vect{x}_2)=\exp{(-[(0.8-0.5)^2+(-0.2-0.4)^2])}=\exp{(-[0.09+0.36])}=\exp(-0.45)\approx 0.6376$.
% Consequently, this pairwise value forms the off-diagonal entry of the resulting kernel matrix.%=(0.8-0.5)^2+(-0.2-0.4)^2

%\begin{figure}[!t]
    %\centering
    %\includegraphics[width=\columnwidth, height=5.0cm, keepaspectratio]{PCA_IoT.png}
    %\caption{An illustration of the data complexity using two PCA components.}
    %\label{fig:data}
%\end{figure}

\section{Methodology} 
\label{sec:methodology}
%\begin{figure*}[!t]
%    \centering
%    \includegraphics[width=\textwidth, height=4.4cm]{Circuit-edit.png}
%    \caption{Schematic diagram of the distinct feature maps used for PQKs: {(a) $\tn{Rot}(\mat{X})$, a single qubit $\mat{R}_{\tn{X}}$ %     rotation, inspired the designing of (e) $\tn{Rot}(\mat{X})^{\tn{en}}$, $\tn{Rot}(\mat{X})$ with a specific pair of entanglement; %    (b) $\tn{Rot}(\mat{X},\mat{Y},\mat{Z})$, rotations along $\mat{R}_{\tn{X}}$, $\mat{R}_{\tn{Y}}$, $\mat{R}_{\tn{Z}}$ gates; (c) %   %      $\tn{IQP}$, standard design, for simplicity, only an $n=3$ qubits circuit is shown; (d) $\tn{Rot}(\mat{X},\mat{Y},\mat{Z})\tn{+CNOT}$, $\tn{Rot}(\mat{X},\mat{Y},\mat{Z})$ with circular entanglement, conceptually formed (f) $\tn{Rot}%%(\mat{X,Y,Z})^{\tn{en}}$, with a specific pair of entanglement.}} %^{\dagger} 
%    \label{fig:circuit}
%\end{figure*}
\subsection{Projected Quantum Kernel Classifier}  
\label{sec:classify}

The IoT input features are encoded into high-dimensional quantum states using $6$-qubit quantum feature-map circuits, where each qubit encodes one input feature. The resulting quantum states are then measured using local Pauli observables to extract information from localized subsystems. Following the subsystem measurement strategies described above, these quantum states are projected into a classical feature space by evaluating the expectation values of the selected local observables. The resulting expectation values are concatenated to form lower-dimensional classical feature vectors, which are then used to compute all pairwise PQK $\kappa_{\tn{PQK}}(\vect{x},\vect{x}')$ values. After all pairwise PQK values are computed, the remaining classification stage is entirely classical. In particular, PQK classification follows the standard SVM-kernel pipeline~\cite{Huang-etal21_1, SchnabelRoth25_1, D’Amore-etal26_1}, with the classical feature map replaced by the PQK pipeline described above. 

Consider the binary training set $\set{D}=\{(\vect{x}_i,y_i)\}_{i=1}^{N}$ with labels $y_i\in\{-1,+1\}$. The values $\kappa_{\tn{PQK}}(\vect{x}_i,\vect{x}_j)$, $i,j\in [N]$, are supplied to a classical soft-margin SVM classifier, which learns the dual coefficients $\vect{\alpha}=(\alpha_1,\ldots,\alpha_N)$, by solving the optimization problem
\begin{IEEEeqnarray*}{rCl}
    \vect{\alpha}^{\star}
    & = &\underset{\bm{\alpha}}\argmin\left\{
    \frac{1}{2}\sum_{i=1}^{N}\sum_{j=1}^{N}\alpha_i\alpha_j y_i y_j\kappa_{\tn{PQK}}(\vect{x}_i,\vect{x}_j)-\sum_{i=1}^{N}\alpha_i\right\}
    \nonumber\\
    && \tn{subject to}\quad
    0 \leq \alpha_i \leq \xi,\quad i\in [N],\quad\sum_{i=1}^{N}\alpha_i y_i=0.
    \label{eq:pqk_svm_dual}
\end{IEEEeqnarray*}
The support vectors are the training samples with $\alpha_i^{\star}>0$. For the classical RBF kernel, the optimization formulation remains unchanged, with $\kappa_{\tn{PQK}}$ replaced by the RBF kernel $\kappa_{\tn{RBF}}=\exp{(-\gamma\enorm{\vect{x}_i-\vect{x}_j}^2)}$ before hyperparameter tuning.

\begin{figure*}[t!]% [htbp!]
\centering
\resizebox{\textwidth}{!}{
\begin{tikzpicture}[
    % --- Custom Styles based on your template ---
    qgate/.style={rectangle, draw=black, thick, fill=white, minimum width=18pt, minimum height=18pt, align=center, rounded corners=2pt, font=\sffamily\scriptsize},
    qdot/.style={circle, fill=black, inner sep=0pt, minimum size=4pt},
    qtarget/.style={circle, draw=black, thick, inner sep=0pt, minimum size=10pt, path picture={
        \draw[thick] (path picture bounding box.north) -- (path picture bounding box.south);
        \draw[thick] (path picture bounding box.west) -- (path picture bounding box.east);
    }},
    meter/.style={rectangle, draw=black, thick, fill=white, minimum width=18pt, minimum height=18pt, rounded corners=1pt, path picture={
        \draw[thick] ([xshift=2.5pt, yshift=5pt]path picture bounding box.south west) arc (180:0:6.5pt);
        \draw[thick, ->, >=stealth] ([xshift=9pt, yshift=5pt]path picture bounding box.south west) -- ([xshift=15pt, yshift=14pt]path picture bounding box.south west);
    }},
    label text/.style={font=\sffamily\color{black}}, %blue!60!cyan
    arrow line/.style={line width=2.5pt, blue!60!cyan, ->, >=stealth}
]
% Define vertical spacing for the qubits
\def\dy{-0.95} 
% --- Block (a): Initial RX Gates ---
\begin{scope}[shift={(0,0)}]
    \foreach \i in {0,...,5} {
        \node[font=\sffamily\scriptsize, left] at (0, \i*\dy) {\i};
        \draw[thick] (0, \i*\dy) -- (1.4, \i*\dy);
        \node[qgate] at (0.8, \i*\dy) {$\mat{R}_{\tn{X}}$};
    }
    \node[label text] at (0.7, 5.9*\dy) {\tn{(a)} $\tn{Rot}(\mat{X})$};
\end{scope}
% --- First Transition Arrow ---
\begin{scope}[shift={(1.6, 2.5*\dy)}]
    \draw[arrow line] (0, 0) -- (0.5, 0);
\end{scope}
% --- Block (e): RX Gates + CNOTs ---
\begin{scope}[shift={(2.3,0)}]
    \foreach \i in {0,...,5} {
        \draw[thick] (0, \i*\dy) -- (2.2, \i*\dy);
        \node[qgate] at (0.6, \i*\dy) {$\mat{R}_{\tn{X}}$};
    }
    % CNOT 1 (q2 to q4)
    \draw[thick] (1.4, 2*\dy) -- (1.4, 4*\dy);
    \node[qdot] at (1.4, 2*\dy) {};
    \node[qtarget] at (1.4, 4*\dy) {};
    % CNOT 2 (q4 to q5)
    \draw[thick] (1.8, 4*\dy) -- (1.8, 5*\dy);
    \node[qdot] at (1.8, 4*\dy) {};
    \node[qtarget] at (1.8, 5*\dy) {};
    
    \node[label text] at (1.0, 5.9*\dy) {\tn{(e)} $\tn{Rot}(\mat{X})^{\tn{en}}$};
\end{scope}

% --- Vertical Dashed Divider ---
\draw[dashed, blue!60!cyan, thick, opacity=0.5] (4.8, 0.5) -- (4.8, 5.5*\dy);
% --- Block (b): RX, RY, RZ Sequence ---
\begin{scope}[shift={(5.1,0)}]
    \foreach \i in {0,...,5} {
        \draw[thick] (0, \i*\dy) -- (2.6, \i*\dy);
        \node[qgate] at (0.5, \i*\dy) {$\mat{R}_{\tn{X}}$};
        \node[qgate] at (1.3, \i*\dy) {$\mat{R}_{\tn{Y}}$};
        \node[qgate] at (2.1, \i*\dy) {$\mat{R}_{\tn{Z}}$};
    }
    \node[label text] at (1.2, 5.9*\dy) {\tn{(b)} $\tn{Rot}(\mat{X},\mat{Y},\mat{Z})$};
\end{scope}
% --- Block (c): Middle Entanglement Block (3 Wires) ---
\begin{scope}[shift={(8, 1.5*\dy)}]
    % --- Background wires (No loop, extended to cover all gates) ---
    \draw[thick] (0, 0) -- (4.9, 0);
    \draw[thick] (0, 1*\dy) -- (4.9, 1*\dy);
    \draw[thick] (0, 2*\dy) -- (4.9, 2*\dy);
    
    % --- Initial Gates ---
    \node[qgate] at (0.5, 0) {H};
    \node[qgate] at (0.5, 1*\dy) {H};
    \node[qgate] at (0.5, 2*\dy) {H};
    
    \node[qgate] at (1.2, 0) {$\mat{R}_{\tn{Z}}$};
    \node[qgate] at (1.2, 1*\dy) {$\mat{R}_{\tn{Z}}$};
    \node[qgate] at (1.2, 2*\dy) {$\mat{R}_{\tn{Z}}$};
    
    % --- ZZ Interactions ---
    % Pair 1: Qubit 1 to Qubit 0
    \draw[thick] (1.8, 0) -- (1.8, 1*\dy); \node[qdot] at (1.8, 1*\dy) {}; \node[qtarget] at (1.8, 0) {};
    \node[qgate] at (2.3, 1*\dy) {$\mat{R}_{\tn{Z}}$}; 
    \draw[thick] (2.8, 0) -- (2.8, 1*\dy); \node[qdot] at (2.8, 1*\dy) {}; \node[qtarget] at (2.8, 0) {};

    % Pair 2: Qubit 2 to Qubit 0
    \draw[thick] (3.3, 0) -- (3.3, 2*\dy); \node[qdot] at (3.3, 2*\dy) {}; \node[qtarget] at (3.3, 0) {};
    \node[qgate] at (4, 2*\dy) {$\mat{R}_{\tn{Z}}$}; 
    \draw[thick] (4.5, 0) -- (4.5, 2*\dy); \node[qdot] at (4.5, 2*\dy) {}; \node[qtarget] at (4.5, 0) {};
    
    % --- Block Label ---
    \node[label text] at (2.5, 3*\dy) {\tn{(c)} $\tn{IQP}$};
\end{scope}
% --- Block (d): Full Rotation & Cascade CNOTs ---
\begin{scope}[shift={(13.1,0)}]
    \foreach \i in {0,...,5} {
        \draw[thick] (0, \i*\dy) -- (5.7, \i*\dy);
        \node[qgate] at (0.5, \i*\dy) {$\mat{R}_{\tn{X}}$};
        \node[qgate] at (1.3, \i*\dy) {$\mat{R}_{\tn{Y}}$};
        \node[qgate] at (2.1, \i*\dy) {$\mat{R}_{\tn{Z}}$};
    }
    % Cascade CNOTs
    \draw[thick] (2.8, 0) -- (2.8, 1*\dy); \node[qdot] at (2.8, 0) {}; \node[qtarget] at (2.8, 1*\dy) {};
    \draw[thick] (3.3, 1*\dy) -- (3.3, 2*\dy); \node[qdot] at (3.3, 1*\dy) {}; \node[qtarget] at (3.3, 2*\dy) {};
    \draw[thick] (3.8, 2*\dy) -- (3.8, 3*\dy); \node[qdot] at (3.8, 2*\dy) {}; \node[qtarget] at (3.8, 3*\dy) {};
    \draw[thick] (4.3, 3*\dy) -- (4.3, 4*\dy); \node[qdot] at (4.3, 3*\dy) {}; \node[qtarget] at (4.3, 4*\dy) {};
    \draw[thick] (4.8, 4*\dy) -- (4.8, 5*\dy); \node[qdot] at (4.8, 4*\dy) {}; \node[qtarget] at (4.8, 5*\dy) {};
    \draw[thick] (5.3, 0) -- (5.3, 5*\dy); \node[qdot] at (5.3, 5*\dy) {}; \node[qtarget] at (5.3, 0) {};
    
    \node[label text] at (2.8, 5.9*\dy) {\tn{(d)} $\tn{Rot}(\mat{X},\mat{Y},\mat{Z})\tn{+CNOT}$};
\end{scope}
% --- Second Transition Arrow ---
\begin{scope}[shift={(18.9, 2.5*\dy)}]
    \draw[arrow line] (0, 0) -- (0.5, 0);
\end{scope}
% --- Block (f): Final Layers & Measurements ---
\begin{scope}[shift={(19.5,0)}]
    \foreach \i in {0,...,5} {
        \draw[thick] (0, \i*\dy) -- (5.2, \i*\dy);
        \node[qgate] at (0.5, \i*\dy) {H};
        \node[qgate] at (1.3, \i*\dy) {$\mat{R}_{\tn{Z}}$};
        \node[qgate] at (2.1, \i*\dy) {$\mat{R}_{\tn{X}}$};
        \node[qgate] at (2.9, \i*\dy) {$\mat{R}_{\tn{Z}}$};
        \node[qgate] at (3.7, 5*\dy) {$\mat{R}_{\tn{Z}}$};
        \node[meter] at (5.2, \i*\dy) {};
    }
    % CNOTs
    \draw[thick] (4.3, 2*\dy) -- (4.3, 5*\dy); \node[qdot] at (4.3, 2*\dy) {}; \node[qtarget] at (4.3, 5*\dy) {};
    \draw[thick] (4.6, 1*\dy) -- (4.6, 4*\dy); \node[qdot] at (4.6, 4*\dy) {}; \node[qtarget] at (4.6, 1*\dy) {};

    \node[label text] at (2.5, 5.9*\dy) {\tn{(f)} $\tn{Rot}(\mat{X,Y,Z})^{\tn{en}}$};
\end{scope}
\end{tikzpicture}
}
\vskip-4mm
\caption{Schematic diagram of the distinct feature maps used for PQKs: {(a) $\tn{Rot}(\mat{X})$, a single qubit $\mat{R}_{\tn{X}}$ rotation, inspired the designing of (e) $\tn{Rot}(\mat{X})^{\tn{en}}$, $\tn{Rot}(\mat{X})$ with a specific pair of entanglement; (b) $\tn{Rot}(\mat{X},\mat{Y},\mat{Z})$,   $\mat{R}_{\tn{X}}$, $\mat{R}_{\tn{Y}}$, $\mat{R}_{\tn{Z}}$ rotations; (c) $\tn{IQP}$, standard design, for simplicity, only an $n=3$ qubits circuit is shown; (d) $\tn{Rot}(\mat{X},\mat{Y},\mat{Z})\tn{+CNOT}$, $\tn{Rot}(\mat{X},\mat{Y},\mat{Z})$ with circular entanglement, conceptually formed (f) $\tn{Rot}(\mat{X,Y,Z})^{\tn{en}}$, with a specific pair of entanglement.}}
\label{fig:circuit}
\end{figure*}
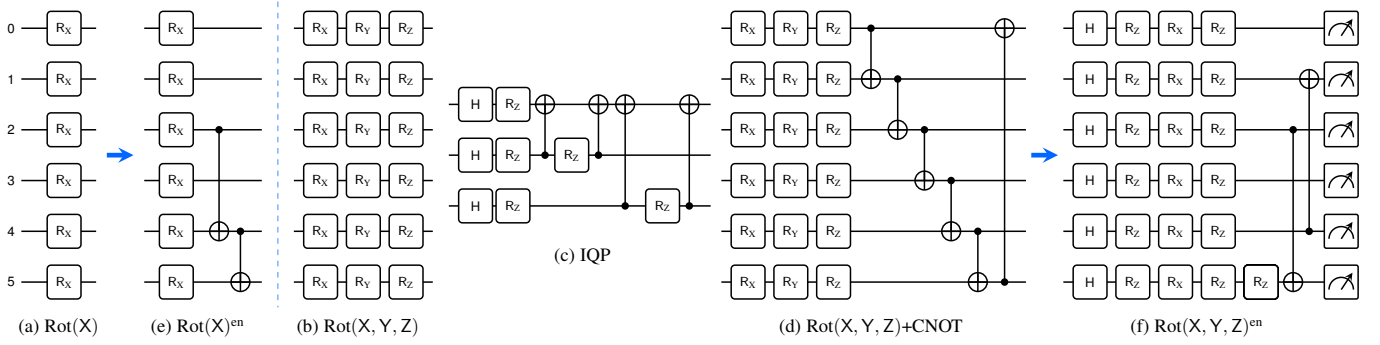
% circuit plot

\subsection{Quantum Feature Map Circuit} 
\label{sec:qfmcircuits}

In QML, quantum feature map circuits are used to implement the data encoding process on a quantum device. In kernel methods, high-order nonlinear feature maps are employed to approximate complex target functions. In the quantum domain, this is widely achieved with instantaneous quantum polynomial ($\tn{IQP}$) designs such as the $\mat{ZZ}$ feature map, which enables nonlinear feature interactions using single-qubit transformations and multi-qubit entanglements~\cite{Havlicek-etal19_1}. Earlier studies have introduced useful encoding designs, properly renamed as $\tn{Rot}(\mat{X})$, a single qubit $\mat{R}_{\tn{X}}$ rotation, and $\tn{IQP}$. Later work also introduced variants of mapping designs, such as $\tn{Rot}(\mat{X,Y,Z})$, $\mat{R}_{\tn{X}}$, $\mat{R}_{\tn{Y}}$, $\mat{R}_{\tn{Z}}$ rotations; and $\tn{Rot}(\mat{X},\mat{Y},\mat{Z})\tn{+CNOT}$, $\tn{Rot}(\mat{X},\mat{Y},\mat{Z})$ with circular entanglement, to develop PQKs for benchmarking and real-world IoT applications~\cite {Huang-etal21_1, SchnabelRoth25_1, D’Amore-etal26_1}. The common idea behind such variants is to introduce a broader class of entanglements, either linear or fully entangled, or additional complexity through additional rotation gates~\cite{D’Amore-etal26_1}. Moreover, a PQK employs distinct measurement strategies described in Sec.~\ref{sec:pqkintro}.

These previous studies support the merit of PQK formulations and their applicability, to some extent, toward developing useful quantum models. However, we observe two major shortcomings: (a) the performance is highly dependent on extensive classical hyperparameter tuning of the RBF kernel parameter $\gamma$ and the SVM regularization parameter $\xi$; and (b) there is a lack of structural explainability in quantum feature-map designs for real-world data. Through this work, we aim to assess the impact of different feature maps without tuning the hyperparameters $\gamma$ and $\xi$, thereby more precisely isolating the representational capabilities of the quantum feature maps. We also propose a structured approach to feature-map design based on selected entanglement pairs. Based on this selection procedure, we propose two new tailor-made feature maps, $\tn{Rot}(\mat{X})^{\tn{en}}$ and $\tn{Rot}(\mat{X,Y,Z})^{\tn{en}}$, for IoT data classification, as illustrated in Fig.~\ref{fig:circuit}~(e) and (f).

The proposed feature maps aim to explore potential empirical advantage by designing entanglement-induced PQKs within our framework. A crucial component of our framework is the intended structural design of the entanglement layer. Rather than fixing an arbitrary entangling pattern, we systematically evaluate pairwise entanglement candidate topologies constructed from all possible two-qubit feature-interaction pairs. We then select the topology that yields the best empirical performance. In this way, the proposed feature maps test whether carefully chosen pairwise entanglement can improve the resulting PQK on real-world IoT data. Specifically, $\tn{Rot}(\mat{X})^{\tn{en}}$ and $\tn{Rot}(\mat{X,Y,Z})^{\tn{en}}$ couple selected qubit pairs through feature-dependent two-qubit interactions, thereby encoding multivariate input-feature correlations into the quantum Hilbert space. Since previous benchmark studies have suggested that \emph{quantumness} via entanglement may not be a decisive performance driver for smaller learning tasks~\cite{BowlesAhmedSchuld24_1sub}, our experiment evaluates whether such selected entanglement structures are beneficial for real-world IoT data.

\subsection{Evaluation Criteria} 
\label{sec:metrics}
To examine the influence of classical hyperparameter tuning, we compare PQKs and the classical RBF baseline before and after tuning. For the untuned PQK baselines, we fix $\xi=1$ and use the data-dependent heuristic $\gamma=\nicefrac{1}{(\ecard{\set{S}}\ecard{\set{P}}\sigma^2)}$, motivated by the variance-scaled bandwidth grid in~\cite[Supplementary information, eq.~(S115)]{Huang-etal21_1}, with the multiplicative factor fixed to one. Here, $\set{S}$ and $\set{P}$ are defined in Sec.~\ref{sec:pqkintro}, and $\sigma^2$ is the empirical variance of the projected features over the training data. For the combined setting, the heuristic uses the $36$-dimensional concatenated training representation. For the untuned classical RBF baseline, we also fix $\xi=1$ and use $\gamma=1/(d\sigma_\textnormal{x}^2)$, where $d=6$ is the input dimension and $\sigma_\textnormal{x}^2$ is the empirical variance of the input features over the training data. This prescribed rule provides a reproducible baseline without grid-search tuning, although the resulting performance comparisons may still depend on the selected baseline hyperparameters.
% \textcolor{red}{The experiments examine intrinsic inductive bias by eliminating extensive tuning of the kernel hyperparameter $\gamma$ and the SVM hyperparameter $\xi$ for both PQKs and RBF. The idea is to isolate the inherent geometric representation of PQKs relative to the baseline RBF under a common training protocol. Doing so, we establish an untuned baseline performance with $\xi=1$ and a data-adaptive heuristic-scale $\gamma = \nicefrac{1}{(\ecard{\set{S}}\ecard{\set{P}} \sigma^2)}$ (cf.~\cite[Supplementary information, eq.~(S115)]{Huang-etal21_1}), where $\set{S}$ and $\set{P}$ are defined in Sec.~\ref{sec:pqkintro}, and $\sigma^2$ represents the empirical variance of the features over the training data. Because the choice of $\gamma$ is data-dependent rather than an arbitrary value, it provides a controlled baseline that isolates representations of kernels.}
For the PQKs only, each feature-map circuit is repeated twice to increase its expressivity and circuit complexity, which may make classical simulation more demanding as the system size increases~\cite{Havlicek-etal19_1, JhaKasabovBhattacharyyaCoylePrasad25_2}. Thereafter, we tune each feature map circuit to quantify its impact and assess the need for tuning. We have used $10$-fold cross-validation, in which $10\%$ of the data is used for testing, with $95\%$ confidence intervals (CIs) to assess the performance of untuned and tuned models, and report test accuracy. The optimal values for the tuning hyperparameters are determined through a grid search over the sets $\xi\in \{0.006, 0.015, 0.03, 0.0625\} \cup \{2^r \mid r=-3,\ldots,10\}$ and $\gamma \in \{0.03, 0.0625\}\cup\{2^r \mid r=-3,\ldots,2\} \cup \{5,7,8\}$~\cite{D’Amore-etal26_1}.

The PQKs performances are evaluated using distinct statistical metrics such as \emph{accuracy}, \emph{F1-score}, and \emph{Matthews correlation coefficient (MCC)}. MCC provides valuable consideration, given that IoT data is slightly imbalanced and reported within the standard range $[-1, +1]$~\cite{ChiccoJurman20_1, JhaKasabovBhattacharyyaCoylePrasad25_2}. In addition to linear (Lin) and RBF kernels, other non-kernel ML models such as linear discriminant analysis (LDA), random forest (RF), AdaBoost (AB), and multilayer perceptron (MLP), are included for a robust and realistic comparison beyond the kernel methods. These non-kernel models are also optimized (tuned) to achieve their best performance. The experiments are conducted in a noiseless quantum PennyLane simulator~\cite{Bergholm-etal18_1sub}. 

\begin{table}[t!]
\caption{IoT data classification evaluation for classical models.} % The untuned scores are highlighted in blue.
\label{tab:rbf}
\centering
\vskip-2mm
\renewcommand{\arraystretch}{1.35}
\begin{tabular}{|c|>{$}c<{$}>{$}c<{$}>{$}c<{$}|}
\hline
\textbf{Models}
& \textbf{Accuracy} & \textbf{F1-score} & \textbf{MCC} \\
\hline
\cellcolor{white}\multirow{1}{*}{$\tn{Lin}$} 
& 0.792\pm0.014 & 0.798\pm0.015 & 0.626\pm0.023 \\
\hline
\rowcolor{blue!0}
\cellcolor{white}\multirow{2}{*}{$\tn{RBF}$}
& \blue 0.826\pm0.017 & \blue 0.820\pm0.021 & \blue 0.670\pm0.033 \\
& 0.893\pm0.008 & 0.881\pm0.009 &  0.785\pm0.017 \\
\hline
\cellcolor{white}\multirow{1}{*}{$\tn{LDA}$}
& 0.792\pm0.012 & 0.798\pm0.013 & 0.627\pm0.017 \\
\hline
\cellcolor{white}\multirow{1}{*}{$\tn{RF}$}
& 0.902\pm0.006 & 0.891\pm0.008 & 0.804\pm0.013 \\
\hline
\cellcolor{white}\multirow{1}{*}{$\tn{AB}$}
& 0.863\pm0.010 & 0.839\pm0.016 & 0.720\pm0.023 \\
\hline
\cellcolor{white}\multirow{1}{*}{$\tn{MLP}$}
& 0.891\pm0.007 & 0.878\pm0.008 & 0.781\pm0.014\\
\hline
\end{tabular}
\end{table}

\section{Numerical Results and Discussion} 
\label{sec:results}

\begin{table*}[t!]
\centering
\caption{IoT data classification accuracy, F1-score, and MCC value for PQK models before and after hyperparameter tuning.} % with three measurement techniques The untuned scores are highlighted in blue.
\label{tab:pqks}
\vskip-2mm
\resizebox{\textwidth}{!}{%
\renewcommand{\arraystretch}{1.35}
\setlength{\tabcolsep}{4pt}
\begin{tabular}{|c|>{$}c<{$}>{$}c<{$}>{$}c<{$}|>{$}c<{$}>{$}c<{$}>{$}c<{$}|>{$}c<{$}>{$}c<{$}>{$}c<{$}|}
\hline
\multirow{2}{*}{\textbf{Models}}
& \multicolumn{3}{c|}{\textbf{\textit{$m=1$}}}
& \multicolumn{3}{c|}{\textbf{\textit{$m=2$}}}
& \multicolumn{3}{c|}{\textbf{\textit{$m \in \{1, 2\}$}}} \\
\cline{2-10}
& \textbf{Accuracy} & \textbf{F1-score} & \textbf{MCC}
& \textbf{Accuracy} & \textbf{F1-score} & \textbf{MCC}
& \textbf{Accuracy} & \textbf{F1-score} & \textbf{MCC} \\
\hline
\rowcolor{blue!0}
\cellcolor{white}\multirow{2}{*}{$\footnotesize \tn{Rot}(\mat{X})$}
& \blue 0.853\pm0.015 & \blue 0.842\pm0.017 & \blue 0.711\pm0.030
& \blue 0.855\pm0.013 & \blue 0.847\pm0.012 & \blue 0.720\pm0.026
& \blue 0.859\pm0.015 & \blue 0.850\pm0.016 & \blue 0.726\pm0.030 \\
& 0.893\pm0.009 & 0.881\pm0.009 & 0.786\pm0.018
& 0.895\pm0.009 & 0.883\pm0.010 & 0.789\pm0.018
& 0.896\pm0.009 & 0.884\pm0.010 & 0.792\pm0.019 \\
\hline
\rowcolor{blue!0}
\cellcolor{white}\multirow{2}{*}{$\footnotesize \tn{Rot}(\mat{X,Y,Z})$} & \blue 0.864\pm0.012 & \blue 0.858\pm0.015 & \blue 0.742\pm0.023 & \blue 0.852\pm0.013 & \blue 0.843\pm0.015 & \blue 0.714\pm0.026 & \blue 0.864\pm0.013 & \blue 0.857\pm0.015 & \blue 0.740\pm0.026 \\
& 0.895\pm0.011 & 0.883\pm0.012 & 0.789\pm0.022
& 0.892\pm0.010 & 0.880\pm0.012 & 0.784\pm0.021
& 0.897\pm0.010 & 0.885\pm0.013 & 0.793\pm0.021 \\
\hline
\rowcolor{blue!0}
\cellcolor{white}\multirow{2}{*}{\centering \shortstack{$\footnotesize \tn{Rot}(\mat{X,Y,Z})$ \\ $\tn{+CNOT}$}} & \blue 0.844\pm0.013 & \blue 0.832\pm0.015 & \blue 0.693\pm0.027 & \blue 0.839\pm0.013 & \blue 0.829\pm0.016 & \blue 0.688\pm0.022 & \blue 0.857\pm0.014 & \blue 0.849\pm0.017 & \blue 0.724\pm0.027 \\
& 0.894\pm0.008 & 0.883\pm0.010 & 0.789\pm0.017
& 0.890\pm0.006 & 0.878\pm0.009 & 0.780\pm0.011
& 0.896\pm0.008 & 0.886\pm0.010 & 0.794\pm0.016 \\
\hline
\rowcolor{blue!0}
\cellcolor{white}\multirow{2}{*}{$\footnotesize \tn{IQP}$} & \blue 0.836\pm0.014 & \blue 0.828\pm0.018 & \blue 0.685\pm0.024 & \blue 0.838\pm0.015 & \blue 0.826\pm0.020 & \blue 0.683\pm0.029 & \blue 0.838\pm0.013 & \blue 0.829\pm0.017 &	\blue 0.687\pm0.023 \\
& 0.872\pm0.008 & 0.856\pm0.011 & 0.741\pm0.017
& 0.868\pm0.008 & 0.854\pm0.011 & 0.736\pm0.017
& 0.873\pm0.011 & 0.862\pm0.013 & 0.750\pm0.021 \\
\hline
\rowcolor{blue!0}
\cellcolor{white}\multirow{2}{*}{$ \footnotesize \tn{Rot}(\mat{X})^{\tn{en}}$} & \blue 0.848\pm0.015 & \blue 0.838\pm0.018 & \blue 0.703\pm0.031 & \blue 0.855\pm0.015 & \blue 0.845\pm0.016 & \blue 0.717\pm0.027 & \blue 0.856\pm0.016 & \blue 0.846\pm0.018 & \blue 0.720\pm0.031 \\
& \mathbf{0.902\pm0.009} & \mathbf{0.891\pm0.010} & \mathbf{0.803\pm0.019}
& \mathbf{0.898\pm0.012} & \mathbf{0.887\pm0.015} & \mathbf{0.796\pm0.024}
& \mathbf{0.903\pm0.011} & \mathbf{0.892\pm0.013} & \mathbf{0.806\pm0.022} \\
\hline
\rowcolor{blue!0}
\cellcolor{white}\multirow{2}{*}{$\footnotesize \tn{Rot}(\mat{X,Y,Z})^{\ast}$} & \blue 0.850\pm0.010 & \blue 0.844\pm0.013 &	\blue 0.716\pm0.019 & \blue 0.854\pm0.012 & \blue 0.849\pm0.016 &	\blue 0.726\pm0.024 & \blue 0.856\pm0.010 & \blue 0.850\pm0.014 & \blue 0.727\pm0.019 \\
& {\small 0.893\pm0.010} & 0.883\pm0.011 & 0.789\pm0.020
& 0.890\pm0.010 & 0.879\pm0.009 & 0.782\pm0.019
& 0.894\pm0.008 & 0.883\pm0.009 & 0.789\pm0.015 \\
\hline
\rowcolor{blue!0}
\cellcolor{white}\multirow{2}{*}{$\footnotesize \tn{Rot}(\mat{X,Y,Z})^{\tn{en}}$} & \mathbf{\blue 0.865\pm0.014} & \mathbf{\blue 0.857\pm0.017} &	\mathbf{\blue 0.740\pm0.028} & \mathbf{\blue 0.867\pm0.015} &	\mathbf{\blue 0.855\pm0.020} &	\mathbf{\blue 0.737\pm0.032} & \mathbf{\blue 0.868\pm0.014} & \mathbf{\blue 0.860\pm0.018} & \mathbf{\blue 0.746\pm0.028} \\
& {\small \mathbf{0.900\pm0.006}} & \mathbf{0.889\pm0.008} & \mathbf{0.801\pm0.012}
& \mathbf{0.896\pm0.010} & \mathbf{0.884\pm0.015} & \mathbf{0.791\pm0.022}
& \mathbf{0.903\pm0.008} & \mathbf{0.892\pm0.010} & \mathbf{0.806\pm0.016} \\
\hline
%\multicolumn{10}{l}{\textcolor{blue!80}{$^{\mathrm{*}}$ Untuned scores.}}
\end{tabular}%
}
\end{table*}

The performance on the IoT data is analyzed from two perspectives: (a) the evaluation of classical models to establish baseline performance, with particular emphasis on the RBF kernel under both untuned and tuned settings, as reported in Table~\ref{tab:rbf}; and (b) the evaluation of different PQKs feature maps with and without hyperparameter tuning, as reported in Table~\ref{tab:pqks}. This structure allows us to separate the effect of classical hyperparameter tuning from the intrinsic representational capability of the feature maps. For both Tables~\ref{tab:rbf} and \ref{tab:pqks}, untuned results are highlighted in blue. 

As shown in Table~\ref{tab:rbf}, the simple linear kernel, Lin, gives the lowest baseline performance, with an accuracy of $0.792\pm0.014$ and an MCC of $0.626\pm0.023$. The LDA model, which is also linear, performs similarly, achieving an accuracy of $0.792\pm0.012$ and an MCC of $0.627\pm0.017$. These results indicate that the classification is not well captured by a purely linear decision boundary.

To examine the role of native inductive bias, we focus on the RBF as the classical nonlinear kernel baseline, since its behavior provides a natural reference point for contextualizing the formulation and potential advantages of PQKs. Table~\ref{tab:rbf} compares the untuned RBF performance with the optimized RBF performance after hyperparameter tuning. Before tuning, the RBF achieves moderate performance, with an accuracy of $0.826\pm0.017$ and an MCC of $0.670\pm0.033$. After grid-search optimization, the performance improves substantially, reaching an accuracy of $0.893\pm0.008$ and an MCC of $0.785\pm0.017$. Overall, the optimized RBF outperforms the Lin baseline by approximately $10$\% in accuracy, $8$\% in F1-score, and $16$\% in MCC, while also reducing the CIs across the metrics. This large improvement shows that the RBF is highly sensitive to hyperparameter tuning.

Among the remaining classical models, AB and MLP perform better than the linear baselines, with AB achieving an accuracy of $0.863\pm0.010$ and an MCC of $0.720\pm0.023$, and MLP achieving an accuracy of $0.891\pm0.007$ and an MCC of $0.781\pm0.014$. The RF model gives the strongest classical performance, with an accuracy of $0.902\pm0.006$, an F1-score of $0.891\pm0.008$, and an MCC of $0.804\pm0.013$. Note that RF is a more complex ensemble model than the kernel methods, as it can capture nonlinear and higher-order feature interactions. Hence, it provides a strong classical benchmark for assessing the empirical competitiveness of the PQKs.

Table~\ref{tab:pqks} reports the performance of six PQK models derived from distinct feature maps. The untuned PQK results already reveal a meaningful native inductive bias. For example, the untuned $\tn{Rot}(\mat{X})$ with $m=1$ achieves an accuracy of $0.853\pm0.015$ and an MCC of $0.711\pm0.030$, outperforming the untuned RBF kernel. Similarly, the untuned $\tn{Rot}(\mat{X},\mat{Y},\mat{Z})$ feature map achieves an accuracy of $0.864\pm0.012$ and an MCC of $0.742\pm0.023$ under $m=1$. These results suggest that the induced quantum feature spaces can provide useful geometric separation even before classical hyperparameter tuning is applied.

After hyperparameter tuning, the standard PQK feature maps become comparable to the tuned RBF baseline. For $\tn{Rot}(\mat{X})$, the best performance is obtained under $m \in \{1, 2\}$, with an accuracy of $0.896\pm0.009$, an F1-score of $0.884\pm0.010$, and an MCC of $0.792\pm0.019$. The $\tn{Rot}(\mat{X},\mat{Y},\mat{Z})$ feature map achieves a similar trend, with its best-tuned performance under $m \in \{1,2\}$ reaching an accuracy of $0.897\pm0.010$ and an MCC of $0.793\pm0.021$. These results indicate that increasing the measurement-based feature representation from individual $m$ values to combined values $m \in \{1,2\}$ yields consistent, although marginal improvements.

% Similar trends can be observed for other PQKs. The untuned $\tn{Rot}(\mat{X,Y,Z})$ performed slightly better than untuned $\tn{Rot}(\mat{X})$ with an accuracy of $0.864 \pm 0.012$ and an MCC of $0.742 \pm 0.023$. After tuning, the accuracy was $0.895 \pm 0.011$ and an MCC was $0.789 \pm 0.022$ when $m=1$, with slightly improved performance for $m \in \{1, 2\}$. Moreover, $\tn{Rot}(\mat{X},\mat{Y},\mat{Z})\tn{+CNOT}$, which uses a set of circular entanglement layers, performed close to $\tn{Rot}(\mat{X,Y,Z})$. The latter has no entanglement layers, implying a case where \emph{quantumness} was not suggested to be useful~\cite{BowlesAhmedSchuld24_1sub}. Furthermore, the $\tn{IQP}$ kernel, despite having dense entanglement, underperformed compared with the other previously mentioned PQKs, with an accuracy of $0.873 \pm 0.011$, and an MCC of $0.750 \pm 0.021$ for $m \in \{1, 2\}$. These results demonstrate that standard literature architectures like $\tn{Rot}(\mat{X},\mat{Y},\mat{Z})\tn{+CNOT}$ and $\tn{IQP}$ may not provide substantial empirical advantages. This corroborates the limitations raised in~\cite{BowlesAhmedSchuld24_1sub, SchnabelRoth25_1}.

The entanglement-based standard feature maps do not automatically lead to better performance. In particular, $\tn{Rot}(\mat{X},\mat{Y},\mat{Z})\tn{+CNOT}$ performs similarly to $\tn{Rot}(\mat{X},\mat{Y},\mat{Z})$ after tuning, while the IQP feature map performs worse compared to the other PQKs, reaching an accuracy of $0.873\pm0.011$ and an MCC of $0.750\pm0.021$ under $m \in \{1,2\}$. This observation suggests that simply adding entanglement or using a dense entangling architecture is not useful in improving PQK performance. Rather, the structure of the entanglement pattern appears to be important.

% Now we discuss the results of the proposed feature maps that prioritize strategic entanglement. The $\tn{Rot}(\mat{X})^{\tn{en}}$, an entangled extension of the baseline $\tn{Rot}(\mat{X})$; and $\tn{Rot}(\mat{X,Y,Z})^{\tn{en}}$ with a carefully designed entanglement layer inspired from other mentioned feature maps. The $\tn{Rot}(\mat{X})^{\tn{en}}$ outperformed existing PQKs with an accuracy of $0.902 \pm 0.009$, F1-score of $0.891 \pm 0.010$, and an MCC of $0.803 \pm 0.019$ for $m=1$, and appeared similar for $m \in \{1, 2\}$ with an accuracy of $0.903 \pm 0.011$ and an MCC of $0.806 \pm 0.022$. The $\tn{Rot}(\mat{X,Y,Z})^{\tn{en}}$ also outperformed the rest of the PQKs and all the classical models provided in Table~\ref{tab:rbf} with substantial empirical advantages with an accuracy of $0.900 \pm 0.006$, F1-score of $0.889 \pm 0.009$, and an MCC of $0.801 \pm 0.012$ for $m=1$, and best performance overall with an accuracy of $0.903 \pm 0.008$, F1-score of $0.892 \pm 0.010$, and an MCC of $0.806 \pm 0.016$ for $m \in \{1, 2\}$.

The proposed feature maps, based on selected pairwise entanglement structures, yield the strongest PQK results. The proposed $\tn{Rot}(\mat{X})^{\tn{en}}$ feature map achieves an accuracy of $0.902\pm0.009$, an F1-score of $0.891\pm0.010$, and an MCC of $0.803\pm0.019$ under $m=1$. Under $m \in \{1,2\}$, it further reaches an accuracy of $0.903\pm0.011$ and an MCC of $0.806\pm0.022$. Similarly, $\tn{Rot}(\mat{X},\mat{Y},\mat{Z})^{\tn{en}}$ achieves competitive performance, with an accuracy of $0.900\pm0.006$, an F1-score of $0.889\pm0.008$, and an MCC of $0.801\pm0.012$ under $m=1$, and its best overall performance under $m \in \{1,2\}$ reaches an accuracy of $0.903\pm0.008$, an F1-score of $0.892\pm0.010$, and an MCC of $0.806\pm0.016$. These results show that the proposed selected-entanglement feature maps outperform the standard PQK designs and are competitive with the strongest classical baseline model, the RF. As an ablation study, we also report numerical results for an unentangled variant of $\tn{Rot}(\mat{X,Y,Z})^{\tn{en}}$, denoted by $\tn{Rot}(\mat{X,Y,Z})^\ast$, to isolate and quantify the performance gains attributable to entanglement.

%\subsection{Sensitivity Analysis}%
% Further quantification is provided to illustrate the hyperparameter tuning via a sensitivity analysis in Fig.~\ref{fig:sensitivity}. We define tuning sensitivity as the performance difference between tuned and untuned scores, as $\Delta_{\textnormal{Acc}}$ for accuracy and $\Delta_{\textnormal{MCC}}$ for MCC. As illustrated in Fig.~\ref{fig:sensitivity}, the RBF exhibited extreme sensitivity to hyperparameters, relying heavily on tuning to achieve optimal performance. In contrast, the PQKs demonstrate notably lower sensitivity. This contrast effectively captures the strong native inductive bias of feature maps within our framework. Overall, our study effectively demonstrates (a) the presence of the strong native inductive bias inherent in PQKs and (b) carefully designed quantum feature maps, such as proposed entanglement-induced designs, have the potential to provide empirical advantages exploiting \emph{quantumness}.

% Sensitivity plot
%\begin{figure}[!t]
%    \centering
%    \includegraphics[width=\columnwidth,height=5.5cm]{IEEE_QAI 2026/Sensitivity.png}
%    \caption{Sensitivity analysis of PQKs and RBF hyperparameters tuning.}
%    \label{fig:sensitivity}
%\end{figure}
% This file was created with matplot2tikz v0.4.2.
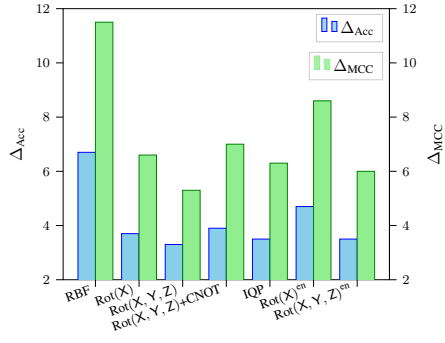
\begin{figure}[!t]
\centering
\begin{adjustbox}{width=0.5\textwidth, height=0.25\textwidth, keepaspectratio}
\begin{tikzpicture}
\definecolor{darkgray176}{RGB}{176,176,176}
\definecolor{green}{RGB}{0,128,0}
\definecolor{lightgray204}{RGB}{204,204,204}
\definecolor{lightgreen}{RGB}{144,238,144}
\definecolor{skyblue}{RGB}{135,206,235}
\begin{axis}[
% width=0.46\textwidth,
% height=0.25\textwidth,
legend cell align={left},
legend style={
  fill opacity=0.8,
  draw opacity=1,
  text opacity=1,
  at={(0.98,0.98)},
  anchor=north east,
  draw=lightgray204
},
tick align=outside,
tick pos=left,
x grid style={darkgray176},
xmin=-0.74, xmax=6.74,
xtick style={color=black},
xtick={0,1,2,3,4,5,6},
xticklabels={\(\displaystyle \tn{RBF}\),\(\displaystyle \tn{Rot}(\mat{X})\),\(\displaystyle \tn{Rot}(\mat{X,Y,Z})\), \(\displaystyle \tn{Rot}(\mat{X,Y,Z})\tn{+CNOT}\),\(\displaystyle \tn{IQP}\),\(\displaystyle \tn{Rot}(\mat{X})^{\tn{en}}\),\(\displaystyle \tn{Rot}(\mat{X,Y,Z})^{\tn{en}}\)},
xticklabel style={font=\footnotesize, rotate=20, anchor=east},
y grid style={darkgray176},
ylabel={\(\displaystyle \Delta_\textnormal{Acc}\)},
ymin=2, ymax=12,
yticklabel style={font=\footnotesize},
ytick style={color=black}
]
\draw[draw=blue,fill=skyblue] (axis cs:-0.4,0) rectangle (axis cs:0,6.7);
\addlegendimage{ybar,ybar legend,draw=blue,fill=skyblue}
\addlegendentry{$\Delta_\textnormal{Acc}$}
\draw[draw=blue,fill=skyblue] (axis cs:0.6,0) rectangle (axis cs:1,3.7);
\draw[draw=blue,fill=skyblue] (axis cs:1.6,0) rectangle (axis cs:2,3.3);
\draw[draw=blue,fill=skyblue] (axis cs:2.6,0) rectangle (axis cs:3,3.9);
\draw[draw=blue,fill=skyblue] (axis cs:3.6,0) rectangle (axis cs:4,3.5);
\draw[draw=blue,fill=skyblue] (axis cs:4.6,0) rectangle (axis cs:5,4.7);
\draw[draw=blue,fill=skyblue] (axis cs:5.6,0) rectangle (axis cs:6,3.5);
\end{axis}
\begin{axis}[
axis y line*=right,
axis x line=none,
legend cell align={left},
legend cell align={left},
legend style={
  fill opacity=0.8,
  draw opacity=1,
  text opacity=1,
  at={(0.98,0.84)},
  anchor=north east,
  draw=lightgray204
},
tick align=outside,
x grid style={darkgray176},
xmin=-0.74, xmax=6.74,
xtick pos=left,
xtick style={color=black},
xtick={0,1,2,3,4,5,6},
xticklabels={\(\displaystyle \tn{RBF}\),\(\displaystyle \tn{Rot}(\mat{X})\),\(\displaystyle \tn{Rot}(\mat{X,Y,Z})\), \(\displaystyle \tn{Rot}(\mat{X,Y,Z})\tn{+CNOT}\),\(\displaystyle \tn{IQP}\),\(\displaystyle \tn{Rot}(\mat{X})^{\tn{en}}\),\(\displaystyle \tn{Rot}(\mat{X,Y,Z})^{\tn{en}}\)},
xticklabel style={font=\footnotesize, rotate=20, anchor=east},
y grid style={darkgray176},
ylabel={\(\displaystyle \Delta_\textnormal{MCC}\)},
ymin=2, ymax=12,
ytick pos=right,
yticklabel style={font=\footnotesize},
ytick style={color=black},
yticklabel style={anchor=west}
]
\draw[draw=green,fill=lightgreen] (axis cs:-2.77555756156289e-17,0) rectangle (axis cs:0.4,11.5);
\addlegendimage{ybar,ybar legend,draw=lightgreen,fill=lightgreen}
\addlegendentry{$\Delta_\textnormal{MCC}$}
\draw[draw=green,fill=lightgreen] (axis cs:1,0) rectangle (axis cs:1.4,6.6);
\draw[draw=green,fill=lightgreen] (axis cs:2,0) rectangle (axis cs:2.4,5.3);
\draw[draw=green,fill=lightgreen] (axis cs:3,0) rectangle (axis cs:3.4,7);
\draw[draw=green,fill=lightgreen] (axis cs:4,0) rectangle (axis cs:4.4,6.3);
\draw[draw=green,fill=lightgreen] (axis cs:5,0) rectangle (axis cs:5.4,8.6);
\draw[draw=green,fill=lightgreen] (axis cs:6,0) rectangle (axis cs:6.4,6);
\end{axis}
\end{tikzpicture}
\end{adjustbox}
\vskip-3mm
\caption{Sensitivity analysis of PQK and RBF hyperparameter tuning.}
\label{fig:sensitivity}
\end{figure}
To quantify the effect of hyperparameter tuning, Fig.~\ref{fig:sensitivity} presents a sensitivity analysis based on the performance differences between the tuned and untuned settings.
% To further quantify the role of hyperparameter tuning, Fig.~\ref{fig:sensitivity} reports a sensitivity analysis based on the difference between \textcolor{red}{a tuned design with cross-validation and an untuned protocol with data-dependent heuristic-scale $\gamma$ performance.} 
We define $\Delta_{\tn{Acc}}$ as the change in accuracy and $\Delta_{\tn{MCC}}$ as the improvement in MCC after tuning. The RBF exhibits the highest tuning sensitivity, indicating that its strong performance depends heavily on grid-search tuning. In contrast, the PQK models generally show lower sensitivity, suggesting that their feature maps provide a stronger native inductive bias before tuning. The proposed entanglement-induced feature maps maintain strong performance while requiring less tuning-induced correction than the RBF. The results support two main observations. First, all untuned PQKs can induce useful native geometric separation in IoT data classification, as evidenced by their performance relative to the untuned RBF baseline. Second, carefully selected pairwise entanglement structures can improve PQK performance more effectively than generic entanglement patterns, such as circular CNOT layers or IQP-style dense entanglement. These findings suggest that the empirical usefulness of entanglement in PQKs depends not only on the presence of entanglement, but also on how the entanglement topology is selected and aligned with the data. % \textcolor{red}{Thus, our initial finding suggests the necessity to evaluate the role of inductive biases in PQK induced quantum circuits. We note that a robust analysis using multiple default hyperparameter choices can be a promising direction.} %towards formulating generalization.

%Overall, our study effectively demonstrates (a) the presence of the strong native inductive bias inherent in PQKs and (b) carefully designed quantum feature maps, such as proposed entanglement-induced designs, have the potential to provide empirical advantages exploiting quantumness.

\section{Conclusion and Future Work}
\label{sec:conclusion}

This study compared six quantum feature maps for IoT classification before and after classical hyperparameter tuning. The proposed circuits, $\tn{Rot}(\mat{X})^{\tn{en}}$ and $\tn{Rot}(\mat{X,Y,Z})^{\tn{en}}$, achieved the strongest PQK performance and were competitive with the strongest classical baseline. Under the prescribed untuned settings, PQKs outperformed the untuned RBF and generally showed smaller tuning-related gains, suggesting useful task-specific inductive biases.
% The study evaluated PQKs derived from six distinct feature maps to analyze the native inductive bias caused by excessive tuning of {\eirik classical hyperparameters}, %$\gamma$ and $\xi$, 
% and assessed the empirical advantages of two proposed entanglement-induced feature map designs on IoT data. The experimental results demonstrated that carefully structured entanglement layers induced quantum circuits, $\tn{Rot}(\mat{X})^{\tn{en}}$ and $\tn{Rot}(\mat{X,Y,Z})^{\tn{en}}$, {\eirik that} can be useful in designing superior PQKs for real-world data. Crucially, the sensitivity analysis revealed that while classical models heavily depend on extensive hyperparameter tuning to reach optimal performance, PQKs can leverage their native inductive bias to separate data geometry significantly better.

This study is limited to heuristic feature-map designs evaluated on a single, relatively small dataset. Future work should assess robustness across multiple baseline hyperparameter choices and datasets, and examine how circuit-induced inductive biases shape the geometry and expressivity of the projected feature space.
% \textcolor{red}{Despite the promising results, the study faces several limitations. Quantum feature map design is heuristic, and entanglement-layer design is evaluated on a single, relatively small dataset. Future work should investigate the optimization requirements of these components and explore more effective feature maps. Future research should also examine how the inductive biases of quantum circuits influence feature-space geometry and expressivity across diverse problems. We anticipate that these directions may uncover feature representations that better leverage the advantages of data encoding and entanglement design, and can help realize the potential of current QML designs.}

%\section*{Acknowledgment}
%The authors acknowledge the financial support provided in part by SURE-AI Centre Grant 357482, Research %Council of Norway.

% \begin{thebibliography}{00}
% \bibitem{b1} 
% \end{thebibliography}
%\[
%%\begin{bmatrix}
%0.8\\
%-0.2
%\end{bmatrix},
%\qquad
%\vect{z}}(\vect{x}_2)=
%\begin{bmatrix}
%0.5\\
%0.4
%\end{bmatrix}.
%\]
%%
\IEEEtriggeratref{24}
%%  The "triggered" command can be changed if desired:
%%
%\IEEEtriggercmd{\enlargethispage{-20cm}}
%%
%%%%%%
%%%%%%%%%%%%%%%%%%%%%%%%%%%%%%%%%%%%%%%%%%%%%%%%%%%%%%%%%%%%%%%%%%%%%
%%% References
% \balance % quick balance the REFs

\bibliographystyle{IEEEtran}
\bibliography{defshort1, biblioHY} % biblioRJ

\end{document}